\let\ifarxiv=\iftrue
\pdfoutput=1

\ifarxiv
\documentclass[12pt,a4paper,final]{article}

\usepackage[nosort]{cite}
\usepackage{amsmath,amssymb}
\usepackage[a4paper,text={490pt,650pt},centering]{geometry}
\usepackage[explicit]{titlesec}

\titleformat{\section}[block]{\normalfont\bfseries\filcenter}{\scshape\thesection}{0.5em}{#1\\\titlerule}
\titlespacing{\section}{5pc}{*3}{*2}[5pc]
\titleformat{\subsection}[runin]{\normalfont\bfseries\filcenter}{\scshape\thesubsection.}{0.5em}{#1}[.---]
\titlespacing{\subsection}{\parindent}{1.5ex plus 0.1ex minus 0.2ex}{0pt}

\makeatletter

\let\@authors\@empty
\let\@email\@empty
\let\@affiliation\@empty
\let\@pdfsubject\@empty
\let\@keywords\@empty
\let\@preprint\@empty

\providecommand{\pdfsubject}[1]{\gdef\@pdfsubject{#1}}
\providecommand{\keywords}[1]{\gdef\@keywords{#1}}
\providecommand{\preprint}[1]{\toks@\expandafter{\@preprint#1\par}\edef\@preprint{\the\toks@}}
\renewcommand{\author}[1]{\ifx\@authors\@empty\toks@\expandafter{#1}\else\toks@\expandafter{\@authors, #1}\fi\edef\@authors{\the\toks@}}
\providecommand{\email}[1]{\ifx\@email\@empty\toks@\expandafter{#1}\else\toks@\expandafter{\@email, #1}\fi\edef\@email{\the\toks@}}
\providecommand{\affiliation}[1]{\gdef\@affiliation{#1}}

\makeatother
\fi

\ifnum\pdfoutput=1\else
\PassOptionsToPackage{hypertex}{hyperref}
\PassOptionsToPackage{draft}{graphicx}
\usepackage{showkeys}
\fi

\usepackage[bookmarks=true,hyperfigures=true]{hyperref}
\usepackage{graphicx}

\allowdisplaybreaks[3]
\providecommand{\hypersetup}[1]{}

\hypersetup{plainpages=false}
\hypersetup{pdfpagemode=UseNone}
\hypersetup{bookmarksnumbered=true}
\hypersetup{pdfstartview=FitH}
\hypersetup{colorlinks=false}

\newcommand{\lrbrk}[1]{\left(#1\right)}

\newcommand{\lint}{\int\limits}
\def\<{\begin{eqnarray}}
\def\>{\end{eqnarray}}

\renewcommand{\[}{\begin{equation}}
\renewcommand{\]}{\end{equation}}

\newcommand{\C}{\mathbb{C}}
\newcommand{\R}{\mathbb{R}}
\newcommand{\nln}{\notag\\}

\begin{document}

\title{Subleading Soft Factor for String Disk Amplitudes}

\author{Burkhard U.\ W.\ Schwab}
 \email{\href{mailto:burkhard\_schwab@brown.edu}{burkhard\_schwab@brown.edu}}

\affiliation{%
Brown University%
\ifarxiv\\\else\ (\fi%
Department of Physics%
\ifarxiv\\\else)\ \fi%
182 Hope St, Providence, RI, 02912%
}

\date{\today}
\pdfsubject{Concerns subleading soft factors in string theory, v1}

\ifarxiv

\makeatletter
\thispagestyle{empty}
\vspace{0.25cm}

\begin{centering}
\begingroup\Large\bf\@title\par\endgroup
\vspace{1cm}

\begingroup\@authors\par\endgroup
\vspace{5mm}

\begingroup\itshape\@affiliation\par\endgroup
\vspace{3mm}

\begingroup\ttfamily\@email\par\endgroup
\vspace{1cm}

\begin{minipage}{17cm}
 \begin{abstract}
We investigate the behavior of superstring disk scattering amplitudes in the presence of a soft external momentum at finite string tension. We prove that there are no $\alpha'$-corrections to the field theory form of the subleading soft factor $S^{(1)}$. At the end of this work, we also comment on the possibility to find the corresponding subleading soft factors in closed string theory using our result and the KLT relations.
 \end{abstract}
\end{minipage}
\vspace{1cm}

\end{centering}

\makeatother


\fi


\makeatletter
\hypersetup{pdftitle={\@title}}%
\hypersetup{pdfsubject={\@pdfsubject}}%
\hypersetup{pdfkeywords={\@keywords}}%
\hypersetup{pdfauthor={\@authors}}%
\makeatother


\section{Introduction and Discussion}
\label{sec:introduction}

The group of large diffeomorphisms of asymptotically flat spacetime in four dimensions first studied by Bondi, van der Burg, Metzner and Sachs -- short BMS -- and known as the BMS group \cite{Bondi:1962px,Sachs:1962wk} has been shown to be responsible for the soft behavior of gravity scattering amplitudes \cite{Cachazo:2014fwa,He:2014laa}. The BMS group is the semi-direct product \[{\rm BMS} = T \ltimes SL(2,\C)\] of the infinite dimensional group of supertranslations $T$ at null infinity $\mathcal{I}$ of asymptotically flat spacetime, and the non-singular transformations of the asymptotic $S^{2}$ which form an $SL(2,\C)$. It was also suggested in \cite{Barnich:2013axa,Barnich:2011mi,Barnich:2009se,Barnich:2011ct} that these transformations of the asymptotic two-sphere could be enhanced to a Virasoro algebra. This Virasoro algebra has become known as the algebra of superrotations. Their impact on the $\mathcal{S}$-matrix of (quantum) gravity has been studied in \cite{Kapec:2014opa}. 

The subleading soft theorem for scattering amplitudes states that in the presence of a soft graviton with momentum $k_{N}=q\to 0$ and polarization tensor $\epsilon_{\mu\nu}$, the $N$-point tree-level gravity amplitude behaves like
\[M_{N}\to \big(S^{(0)}_{\rm g} + S^{(1)}_{\rm g} + S^{(2)}_{\rm g}\big) M_{N-1}.\] Here $S^{(0)}_{\rm g}$ is Weinberg's soft graviton factor \cite{Weinberg:1965nx,Weinberg:1964ew}, and $S^{(1)}_{\rm g}$ and $S^{(2)}_{\rm g}$ are the new subleading terms \[S^{(0)}_{\rm g} = \sum_{i=1}^{N-1}\frac{\epsilon_{\mu\nu}k_{i}^{\mu}k_{i}^{\nu}}{q.k_{i}},\quad S^{(1)}_{\rm g} = \sum_{i=1}^{N-1}\frac{\epsilon_{\mu\nu}k_{i}^{\mu}(q_{\rho}J^{\rho\nu}_{i})}{q.k_{i}},\quad S^{(2)}_{\rm g} = \sum_{i=1}^{N-1}\frac{\epsilon_{\mu\nu}(q_{\lambda}J^{\lambda\mu}_{i})(q_{\rho}J^{\rho\nu}_{i})}{q.k_{i}}.\label{eq:gravsoft}\] The subleading factors depend on the angular momentum operators $J^{\mu\nu}_{i} = L^{\mu\nu}_{i} + S^{\mu\nu}_{i}$ where $L^{\mu\nu}_{i}$ is the orbital angular momentum operator of particle $i$ and $S^{\mu\nu}_{i}$ is the spin contribution. Gauge invariance of Weinberg's soft graviton factor $S^{(0)}_{\rm g}$ follows from conservation of momentum, while the gauge invariance of the subleading term $S^{(1)}_{\rm g}$ follows from global conservation of angular momentum. The gauge invariance of $S^{(2)}_{\rm g}$ follows from the antisymmetry of $J^{\mu\nu}_{i}$.

It was realized by Casali \cite{Casali:2014xpa} using similar methods as have been used by Cachazo and Strominger \cite{Cachazo:2014fwa}, that there are not only universal subleading soft factors in gravity in four dimensions, but also in Yang-Mills theory \footnote{This subleading soft factor was already known for quite some time \cite{Low:1958sn,Burnett:1967km}. In fact, the subleading soft factor in gravity had been derived before, too \cite{Gross:1968in,White:2011yy}. However, the symmetry principle behind these factors and their universality were unknown at the time. We thank Andrew Larkoski for pointing out these references.}. In Yang-Mills theory, the soft behavior of a color-ordered $N$-point scattering amplitude is given by  \[A_{N}(1,\ldots,N-1,q)\to \big(S^{(0)}_{\rm YM} + S^{(1)}_{\rm YM}\big)A_{N-1}(1,\ldots, N-1)\] where $S^{(0)}_{\rm YM}$ is the soft gluon factor and $S^{(1)}_{\rm YM}$ is the subleading contribution \[S^{(0)}_{\rm YM} = \frac{\epsilon.k_{1}}{q.k_{1}} - \frac{\epsilon.k_{N-1}}{q.k_{N-1}},\qquad S^{(1)}_{\rm YM} = \frac{\epsilon_{\mu}q_{\nu}J^{\mu\nu}_{1}}{q.k_{1}} - \frac{\epsilon_{\mu}q_{\nu}J^{\mu\nu}_{N-1}}{q.k_{N-1}}.\label{eq:weinbgluon}\] Here $\epsilon_{\mu}$ is the polarization vector of the soft particle with momentum $q$. In Yang-Mills theory, gauge invariance of the soft factors follows from the antisymmetry of $J^{\mu\nu}_{i}$. It has been shown that their form is constrained by conformal symmetry \cite{Larkoski:2014hta}.  Furthermore, these subleading factors are also universal in any dimension \cite{Schwab:2014xua}\footnote{For a more detailed calculation of the same result, see \cite{Afkhami-Jeddi:2014fia}.}. The last property has been linked to diffeomorphisms of ambitwistor space in any dimension and studied using ambitwistor string models by Adamo \emph{et al.} and Geyer \emph{et al.} \cite{Adamo:2014yya,Geyer:2014lca}. 

Further exploring the parameter space, the behavior of subleading soft factors in field theory for loop level integrals (see, e.g., the works \cite{Dixon:2009ur,Dixon:2008gr,Bern:1998sv,Dunbar:2012aj}) has been investigated in \cite{Bern:2014oka,He:2014bga}. In these papers it was shown that the subleading soft factors receive corrections due to discontinuities in the loop results. However, Cachazo and Yuan \cite{Cachazo:2014dia} suggested that a modification of the usual procedure of taking the soft limit might yield uncorrected subleading soft factors\footnote{See Bern \emph{et al.} \cite{Bern:1998sv} for a comment on this modification.}.
\newline

The purpose of this paper is to explore another parameter. We examine whether there are $\alpha'$-contributions to the subleading soft factor in color-ordered open string disk amplitudes $\mathcal{A}_{N}$. Since the infinite string tension limit $\alpha'\to 0$ is known to reproduce field theory amplitudes, we would expect that such corrections may first appear at $\mathcal{O}(\alpha')$ in the $\alpha'$-expansion. We will show that this intuition is correct. In fact, we can prove that tree-level string scattering amplitudes show the same behavior in the limit of a soft string as field theory amplitudes do.

Let us make a rough analysis of the possible form of corrections. The soft factors in field theory are classified by their scaling in $q$, i.e., the leading gauge theory soft factor scales like $q^{-1}$ while the subleading factor scales like $q^{0}$. This pattern obviously has to be preserved in string theory. Furthermore, the soft factors have mass dimension zero. A priori, nothing prevents us from allowing ``long-range'' corrections involving additional polarization vectors and hard momenta apart from the appearing adjacent momenta $k_{1}$ and $k_{N-1}$ in \eqref{eq:weinbgluon}. This however could spoil the universality of the soft factor in string theory. Further excluding the appearance of any additional poles in the momenta leads us to expect that any correction to the soft factors might come as polynomials in the dimensionless Mandelstam variables $s_{ij}$. We will see that no such corrections appear. 

The calculations below have become feasible due to developments since the year 2000, following Berkovits's work \cite{Berkovits:2000fe} and the developments in \cite{Mafra:2010jq}. Since then there has been tremendous progress \cite{Mafra:2011nv,Mafra:2011nw} in determining the tree-level string theory amplitudes for arbitrary dimension, compactification and any amount of supersymmetry. Although there has been much more progress in the field we will only need a very small portion of this body of work (see e.g.  \cite{Stieberger:2009hq,Stieberger:2013nha,Stieberger:2013hza,Schlotterer:2012ny,Stieberger:2014hba,Stieberger:2013wea,Broedel:2013aza,Broedel:2013tta} for progress in determining the $\alpha'$-expansion of arbitrary disk amplitudes as well as \cite{Bjerrum-Bohr:2014qwa} for a link between string scattering amplitudes and the Cachazo-He-Yuan (CHY) integral formula for gauge and gravity amplitudes in arbitrary dimension from scattering equations \cite{Cachazo:2013iea,Cachazo:2013hca}).
\newline

In lieu of a table of contents we shall give a description of the organization of the article here. In sec.~\ref{sec:general} we describe the conventions and give necessary definitions for the calculation of the subleading soft limit for disk scattering amplitudes to keep the work reasonably self-contained. The core of this calculation can be found in sec.~\ref{sec:interg-appr-new} where we will use an approximation of the Euler integrals appearing in the string scattering amplitudes which is valid in the soft limit to subleading order. Finally, in sec.~\ref{sec:closed-string-klt} we will remark on the connection between disk string scattering amplitudes and closed string scattering amplitudes and suggest ways to deduce the subleading soft behavior of closed string amplitudes for the soft limits of open strings.

\section{Conventions and general definitions}
\label{sec:general}

Superstring disk scattering amplitudes with $N$ external particles, valid for any dimension $D$, any compactification and any amount of supersymmetry can be given in a surprisingly compact form \cite{Mafra:2011nv,Mafra:2011nw} 
\[\mathcal{A}(1,\ldots,N) = \sum_{\sigma\in S_{N-3}} A_{\rm YM}(1,2_{\sigma},\ldots,(N-2)_{\sigma},N-1,N) F^{\sigma}_{(1,\ldots, N)}(\alpha').\label{eq:ssmform}\]
We follow the conventions given in the references. In the expression we denoted by $A_{\rm YM}$ the basis of $(N-3)!$ color-ordered (super) Yang-Mills amplitudes and by $F^{\sigma}(\alpha')$ the \emph{generalized Euler integrals} which carry the full $\alpha'$-dependence of the string amplitudes. The subscript $\sigma$ on $i_{\sigma}$ denotes the action of the permutation $\sigma\in S_{N-3}$ on the label $i\in (2,\ldots,N-2)$.  Finally, the specific color order of the string scattering amplitude is indicated by the subscript $(1,\ldots, N)$ on $F^{\sigma}_{(1,\ldots, N)}$. The order of this label corresponds to the order of the insertion points of the vertex operators on the boundary of the disk. The functions $F^{\sigma}$ are iterated integrals 
\[F^{(2,3,\ldots,N-2)}_{(1,\ldots,N)}(\{s_{ij}\},\alpha') = (-1)^{N-3} \lint_{z_{i}<z_{i+1}}\prod_{j=2}^{N-2}dz_{j}\ \lrbrk{\prod_{1\leq k<m\leq N-1} |z_{km}|^{s_{km}}} \left\{\prod_{k=2}^{N-2}\sum_{m=1}^{k-1}\frac{s_{mk}}{z_{mk}}\right\}\label{eq:geneuler}\] with the integration regions bounded by using the $SL(2,\R)$-invariance of the disk amplitudes to choose $z_{1}=0$, $z_{N-1}=1$ and $z_{N}=\infty$. More precisely, the notation above indicates that
\[\lint_{z_{i}<z_{i+1}}\prod_{j=2}^{N-2}dz_{j} = \lint_{0}^{1}dz_{2}\lint_{z_{2}}^{1}dz_{3}\cdots\lint_{z_{N-3}}^{1}dz_{N-2}.\]
Different color orders are achieved by changing the order of integrations in the equation above while keeping the integrand fixed. The action of the permutation $\sigma$ (the superscript label) is limited to an action on the curly bracket in \eqref{eq:geneuler}. As is customarily done, we have chosen to hide the $\alpha'$-dependence in the Mandelstam variables \[s_{ij} = \alpha'(k_{i}+k_{j})^{2} = 2\alpha' k_{i}.k_{j}\] where the second equality follows from the gluon momenta being massless and on-shell $k_{i}^{2}=0$. Also, there is no reference to any particular helicity choices, so the results given in the following are true for any choice of helicity structure for the string scattering amplitude.

The soft expansion to leading order has already been done in \cite{Mafra:2011nv,Mafra:2011nw} where it was shown that the behavior of the leading part of the soft theorem is the same as for field theory amplitudes. More precisely, taking an amplitude $\mathcal{A}(1,\ldots,N-2,q,N-1,N)$ with $N+1$ particles and choosing the momentum $k_{q}=q\to 0$ to be soft, one can show that \[\mathcal{A}(1,\ldots,N-2,q,N-1,N) \to \lrbrk{\frac{\epsilon.k_{N-1}}{q.k_{N-1}} - \frac{\epsilon.k_{N-2}}{q.k_{N-2}}}\mathcal{A}(1,\ldots,N)\label{eq:knownfactor}\] where the factor in the bracket is Weinberg's soft gluon factor \eqref{eq:weinbgluon} and $\mathcal{A}(1,\ldots, N)$ is the $N$ particle string disk scattering amplitude. In the following we will make use of this result and its derivation (which can be found in \cite{Mafra:2011nw}).

\section{Soft limit for disk scattering amplitudes}
\label{sec:interg-appr-new}

In this section we will investigate the subleading soft limit of the functions $F^{\sigma}(\alpha')$ and show that the subleading soft factor for tree-level string scattering amplitudes receives no $\alpha'$-corrections relative to the subleading soft factor in field theory. We will first evaluate the appropriate integral in the functions $F^{\sigma}$ in the soft limit in ssec.~\ref{sec:integr-appr-gener}. We will then reorganize the results to show that we get the expected result in ssec.~\ref{sec:reass-subl-fact}. Finally, we will give two low point examples in ssec.s \ref{sec:soft-expansion-4pt} and \ref{sec:soft-expansion-5pt} to illustrate the result.

\subsection{Approximation for Euler integrals}
\label{sec:integr-appr-gener}

We will use an integral approximation\footnote{We would like to thank Steven Avery for pointing out the method.} of \eqref{eq:geneuler} to show that the subleading term in the soft expansion of string scattering amplitudes is in fact given by the field theory factor \[S^{(1)} = \frac{\epsilon_{\mu}q_{\nu}J^{\mu\nu}_{s+1}}{q.k_{s+1}} - \frac{\epsilon_{\mu}q_{\nu}J^{\mu\nu}_{s-1}}{q.k_{s-1}}\] where $s+1$ and $s-1$ denote the particles adjacent to the soft particle in a given color-ordering of the amplitude and $\epsilon_{\mu}$ and $q_{\mu}$ are the polarization vector and momentum of the soft particle. The full disk amplitude is given by \eqref{eq:ssmform}
\[\mathcal{A}_{N} = \sum_{\sigma\in S_{N-3}}A_{\rm YM}(1,2_{\sigma},\ldots,(N-2)_{\sigma},N-1,N)F^{\sigma}_{(12\ldots N)}(\alpha').\label{eq:diskam}\]
Since we know the soft limit of the field theory amplitudes, we will concentrate on the content of the Euler integrals $F^{\sigma}(\alpha')$. We will take the soft limit of a particle $q$ inserted between particle $(N-2)$ and $(N-1)$ in the $(N+1)$-particle amplitude. The question we want to ask is therefore
\[\mathcal{A}(1,2,\ldots,N-2,q,N-1,N) \overset{?}{\to}(S^{(0)} + S^{(1)})\mathcal{A}(1,2,\ldots,N-2,N-1,N).\]

The interesting part of \eqref{eq:geneuler} is the innermost integral which is over $z_{q}$ here. Let us call it $I^{(N-2)}_{q}$ and concentrate on the group of permutations $\sigma_{i}(2,3,\ldots,N-2,q)$ which preserve the order of $(2,3,\ldots,N-2)$ and move\footnote{Notice that $i$ therefore runs from $1$ to $(N-2)$.} $q$ \[\sigma_{i}(2,3,\ldots,N-2,q)= (2,3,\ldots,i,q,i+1,\ldots,N-2).\] All these permutations will lead to the same function $A_{\rm YM}(1,2,\ldots,N)F^{(23\ldots,N-2)}(\alpha')$ in the soft limit with various prefactors. At the end of the calculation we will have to consider their sum. This is done in the following subsection ssec.~\ref{sec:reass-subl-fact}. We also need to remember that -- strictly speaking -- we are performing all of the following operations under the $N-3$ remaining integral signs in $F^{(23\ldots N-2)}_{(1\ldots N)}(\alpha')$. In the following section we will indicate this by using the notation 
\begin{align}
& F^{(2,\ldots,N-2,q)}_{(1\ldots N+1)}(\alpha') = \tilde{F}^{(2,\ldots,N-2)}(\alpha')\star I^{(N-2)}_{q}(\{z_{i}\}_{i\neq q},\{s_{ij}\})\nln 
&\qquad\qquad:= (-1)^{N-3}\lint_{z_{i}<z_{i+1}} \prod_{j}dz_{j}\ \lrbrk{\prod_{1\leq k<m\leq N-1} |z_{km}|^{s_{km}}} \left\{\prod_{k=2}^{N-2}\sum_{m=1}^{k-1}\frac{s_{mk}}{z_{mk}}\right\} I^{(N-2)}_{q}
\end{align}
Here $I^{(N-2)}_{q}$ is part of a function $I_{q}$ which we will define for convenience. It is given by the sum over the contributions of all order-preserving permutations $\sigma_{i}$ to the same subamplitude after the soft limit\footnote{The coefficients $c_{i}$ are bookkeeping devices. They will be replaced by the field theory soft factors $S^{(0)}_{\rm YM}$ in ssec.~\ref{sec:reass-subl-fact}.\label{fn:artificial}}
\[I_{q} = \sum_{i=1}^{N-2}c_{i}I_{q}^{(i)}= - \sum_{i=1}^{N-2}c_{i}\lint_{z_{N-2}}^{1} dz_{q}\ \prod_{j=1}^{N-1} |z_{jq}|^{s_{jq}}\sum_{m=1}^{i}\frac{s_{mq}}{z_{mq}}\lrbrk{\prod_{k=i+1}^{N-2}\sum_{m=1}^{k-1}\frac{s_{mk}}{z_{mk}}}.\] 
To calculate the integral in the soft limit $q\to 0$ we find it useful to consider the two cases $i=N-2$ and otherwise separately. First, take $I_{q}^{(N-2)}$ and break up the sum over $m$ into $m=N-3$ and the rest, i.e.,
\[I_{q}^{(N-2)} = -\lint_{z_{N-2}}^{1}dz_{q} \prod_{i}|z_{iq}|^{s_{iq}}\lrbrk{\sum_{m=1}^{N-3}\frac{s_{mq}}{z_{mq}} + \frac{s_{N-2,q}}{z_{N-2,q}}}.\] The first term (proportional to the sum over $m$) is finite in the limit $k_{q}=q\to 0$, the product goes to $1$ and we can solely keep the leading term as the next term is $\mathcal{O}(q^{2})$. The result is
\[\sum_{m=1}^{N-3}s_{mq}(\log z_{N-1,m} - \log z_{N-2,m}).\] The second part of the integral is slightly harder since the integral diverges when the soft limit is taken before the integration over $z_{q}$. It is possible to calculate the leading and subleading term of this integral by introducing a regulating parameter $\delta$ prior to the approximation of the integral. To do so, break up the integration region \[\lint_{z_{N-2}}^{z_{N-1}} = \lint_{z_{N-2}}^{z_{N-2}+\delta} + \lint_{z_{N-2}+\delta}^{z_{N-1}}\] with $\delta\ll 1$, $z_{N-1}=1$ and approximate the integral in the following way. First, examine the pole at $N-2$
\begin{align}
  \label{eq:firstint}
  -\lint_{z_{N-2}}^{z_{N-2}+\delta}dz_{q} \prod_{i}|z_{iq}|^{s_{iq}}\frac{s_{N-2,q}}{z_{N-2,q}} &=  s_{N-2,q}\lint_{0}^{\delta}dz \prod_{i\neq N-2}|z_{i,N-2}-z|^{s_{iq}}z^{-1+s_{N-2,q}}\nln
&\approx \prod_{i\neq N-2}|z_{i,N-2}|^{s_{iq}}|\delta|^{s_{N-2,q}}
\end{align}
where for the first equality a shift of the integration variable $z_{q}\to z_{q} + z_{N-2}= z$ was performed. Since $z\ll 1$ in the integration region we can drop it everywhere except for $z^{s_{N-2,q}-1}$. The resulting integral is performed easily and yields \eqref{eq:firstint}. The integral over the second region is finite in the soft limit $q\to 0$ so we immediately find
\[-\lint_{z_{N-2}+\delta}^{z_{N-1}}dz_{q}\prod_{i}|z_{iq}|^{s_{iq}}\frac{s_{N-2,q}}{z_{N-2,q}} = s_{N-2,q}\log\lrbrk{\frac{z_{N-1}-z_{N-2}}{\delta}}.\]

Since in the soft limit $s_{iq}\ll 1$ and $\delta\ll 1$, we find that \eqref{eq:firstint} is given by the approximation
\[1 + s_{N-2,q}\log(\delta) +\sum_{i\neq N-2}^{N-1}s_{i,N-2}\log(z_{i,N-3}).\] Notice that treating the pole in $z_{N-2}$ correctly was crucial or we would have ended up with a divergent integral. If we add the two results, the regulation parameter $\delta$ conveniently drops out to next-to-leading order in $q$ and the sums telescope. We find that
\[I_{q}^{(N-2)}\approx 1 + \sum_{i=1}^{N-2} s_{iq}\log z_{N-1,i} + s_{N-1,q}\log z_{N-1,N-2}.\label{eq:approx}\]
The rest of the integral $I_{q}$ follows in a similar way. We have
\[\sum_{i=1}^{N-3}c_{i}I_{q}^{(i)} = - \sum_{i=1}^{N-3}c_{i}\int_{z_{N-2}}^{z_{N-1}}dz_{q}\prod_{j=1}^{N-1}|z_{jq}|^{s_{jq}} \lrbrk{\sum_{n=1}^{i}\frac{s_{nq}}{z_{nq}}}\lrbrk{\prod_{k=i+1}^{N-2}\sum_{m=1}^{k-1} \frac{s_{mk}}{z_{mk}}}.\] The sum over $m$ as well as the product over $k$ also contain the soft particle $q$. At first sight, the expression looks daunting, but there are actually only very few terms that contribute to the leading and next-to-leading order in the integral approximation. 

However, to check that we treat the pole in $z_{N-2}$ correctly, it is necessary to calculate the integral to next-to-next-to-leading order -- this is where the pole appears and, luckily, cancels out for all $i$. The leading term can be extracted from the integral by keeping only the momentum $q$ in the sum over $n$ and taking the product over $j$ to $1$. Then
\[-\sum_{i=1}^{N-3}c_{i}\int_{z_{N-2}}^{z_{N-1}}dz_{q}\lrbrk{\sum_{n=1}^{i}\frac{s_{nq}}{z_{nq}}}\lrbrk{\prod_{k=i+1}^{N-2}{\sum_{m=1}^{k-1}}' \frac{s_{mk}}{z_{mk}}} = \sum_{i=1}^{N-3}c_{i}\lrbrk{\prod_{k=i+1}^{N-2}{\sum_{m=1}^{k-1}}' \frac{s_{mk}}{z_{mk}}}\sum_{m=1}^{i}s_{mq}\log \frac{z_{m,N-1}}{z_{m,N-2}}\label{eq:result2}\] where the prime on the sum indicates that the soft particle is now omitted. The remaining part of the subleading contributions from this integral are found in 
\[-\sum_{i=1}^{N-3}c_{i}\int_{z_{N-2}}^{z_{N-1}}dz_{q}\prod_{j=1}^{N-1}|z_{jq}|^{s_{jq}}\lrbrk{\sum_{n=1}^{i}\frac{s_{nq}}{z_{nq}}}\lrbrk{\prod_{k=i+1}^{N-2}{\sum_{m=1}^{k-1}}' \frac{s_{mk}}{z_{mk}}}\frac{s_{q,N-2}}{z_{q,N-2}}.\] Once again, there is an issue if we take the soft limit before completing the integral due to the pole at $z_{N-2}$. However, the calculation is very similar to the one presented above (regulation and subsequent cancellation of the pole) and we shall only present the result of it, which is
\[\sum_{i=1}^{N-3}c_{i}\lrbrk{\sum_{n=1}^{i}\frac{s_{nq}}{z_{n,N-2}}}\lrbrk{\prod_{k=i+1}^{N-2}{\sum_{m=1}^{k-1}}' \frac{s_{mk}}{z_{mk}}}\label{eq:result3}\] to leading order in $q$. We have now all the pieces of the puzzle in our hands and can proceed to reassemble them into something meaningful.

\subsection{Reassembling the subleading factor}
\label{sec:reass-subl-fact}

Since the subamplitudes $A_{\rm YM}$ are ordinary gauge field theory amplitudes in arbitrary dimensions, we know \cite{Casali:2014xpa,Schwab:2014xua} that \[A_{{\rm YM},N}\to \big(S^{(0)}_{\rm YM} + S^{(1)}_{\rm YM}\big)A_{{\rm YM},N-1}\] in the soft limit with the soft factors as given in \eqref{eq:weinbgluon}. Let us see how the contributions of the field theory amplitudes and the $F^{\sigma}$ functions we calculated above reassemble in the soft limit of the string scattering amplitude.
First, we take, e.g., $A_{\rm YM}(1,2,\ldots,N-2,q,N-1,N)F^{(23\ldots N-2 q)}_{(1\ldots N+1)}(\alpha')$ which is a part of \eqref{eq:diskam}. In the soft limit this part goes to
\[\big(S^{(0)}_{\rm YM} + S^{(1)}_{\rm YM}\big)A_{\rm YM}(1,\ldots,N)F^{(23\ldots N-2)}_{(1\ldots N)}(\alpha')\star\lrbrk{1 + \sum_{m=1}^{N-2}s_{mq}\log z_{N-1,m} + s_{N-1,q}\log z_{N-1,N-2}}\] where -- as we already remarked in the last subsection -- the notation $\star$ indicates that the bracket that follows is considered to be under the $N-3$ iterated integrals of the function\footnote{Notice that $F^{\sigma}$ is quite literally the Euler integral appearing in $\mathcal{A}_{N}$. Compare this with \eqref{eq:subres} where a different function appears.} $F^{(2\ldots N-2)}_{(1\ldots N)}$. The leading term $1$ only contributes to the field theory limit -- which we know already -- so we can ignore it for now.  Proceed by pulling $S^{(0)}_{\rm YM}$ (which is a multiplication operator) through $A_{\rm YM}$ and the $N-3$ integrals of $F^{(2\ldots N-3)}_{(1\ldots N)}$. A simple inspection reveals that the combination \[R_{1} =S^{(0)}_{\rm YM}\lrbrk{\sum_{m=1}^{N-2}s_{mq}\log z_{N-1,m} + s_{N-1,q}\log z_{N-1,N-2}}\] is $\mathcal{O}(q^{0})$. Thus this combination is of the same order in $q$ as the derivative operator $S^{(1)}_{\rm YM}$ acting on the field theory amplitude $S^{(1)}_{\rm YM}A_{\rm YM}$. Thus this term is a contribution to the subleading soft factor. 

We can now inspect all the other permutations corresponding to the same subamplitude $A_{\rm YM,N}$ after the soft limit and perform the same manipulation. This corresponds to restoring Weinberg's soft factors in the function $I_{q}$ for every term in the sum over $i$ by replacing the coefficients $c_{i}\to S^{(0)}_{{\rm YM},i+1}$. It follows that the subleading result for this particular ordering of $(2,3,\ldots,N-2)$ is given by \[(S^{(1)}_{\rm YM}A_{{\rm YM},N})F^{(2\ldots N-2)} + A_{{\rm YM},N}\tilde{F}^{(2\ldots N-2)}\star R\label{eq:subres}\] where we denoted \[R = (R_{1} + R_{2})\sum_{m=1}^{N-3}\frac{s_{m,N-2}}{z_{m,N-2}} + R_{3}\]
with
\begin{align}
R_{2}&=\sum_{i=1}^{N-3}S^{(0)}_{i+1}\sum_{m=1}^{i}\lrbrk{\log z_{m,N-2} - \log z_{m,N-1}}\\
R_{3} &= -\sum_{i=1}^{N-3}S^{(0)}_{i+1}\sum_{m=1}^{i}\frac{s_{mq}}{z_{m,N-2}}.
\end{align}
We want to emphasize that $F^{(2\ldots N-2)}_{(1\ldots N)}=\tilde{F}^{(2\ldots N-2)}_{(1\ldots N)}\star \sum_{m=1}^{N-3}\frac{s_{m,N-2}}{z_{m,N-2}}$.
The first term in \eqref{eq:subres} follows from the same telescoping property which is necessary to show that the string theory amplitudes have the correct leading soft behavior \cite{Mafra:2011nw}.
In the last equations, $S^{(0)}_{i+1}$ is the soft factor corresponding to the soft particle inserted between $i$ and $i+1$. Of course, this is not the simplest form of the function $R$. Using various telescoping properties of the appearing sums, we can write the result as
\begin{align}
R &= \Bigg(\frac{\epsilon.k_{N-1}}{q.k_{N-1}}\sum_{m=1}^{N-2}s_{mq}\log z_{m,N-1} -\sum_{i=1}^{N-2}\frac{\epsilon.k_{i}}{q.k_{i}}s_{iq}\log z_{i,N-1}\nln
 &\qquad -\frac{\epsilon.k_{N-2}}{q.k_{N-2}}\sum_{\overset{m=1}{m\neq N-2}}^{N-1}s_{mq}\log z_{m,N-2} +\sum_{\overset{i=1}{i\neq N-2}}^{N-1}\frac{\epsilon.k_{i}}{q.k_{i}}s_{iq}\log z_{i,N-2}\Bigg)\sum_{m=1}^{N-3}\frac{s_{m,N-2}}{z_{m,N-2}}\nln
 &\qquad - \frac{\epsilon.k_{N-2}}{q.k_{N-2}}\sum_{m=1}^{N-3}\frac{s_{mq}}{z_{m,N-2}} + \sum_{i=1}^{N-3}\frac{\epsilon.k_{i}}{q.k_{i}}\frac{s_{iq}}{z_{i, N-2}}.
\end{align}
This is, amazingly, exactly the contribution we would expect from the operator \[S^{(1)} = \frac{\epsilon_{\mu}q_{\nu}J^{\mu\nu}_{N-1}}{q.k_{N-1}}-\frac{\epsilon_{\mu}q_{\nu}J^{\mu\nu}_{N-2}}{q.k_{N-2}}\] acting on $F^{(23\ldots N-2)}_{(1\ldots N)}(\alpha')$. We can therefore see that for this particular ordering of the labels $(2,\ldots,N-2)$ the subleading contribution to the soft theorem is just \[S^{(1)}(A_{{\rm YM},N}F^{(2\ldots N-2)}_{(1\ldots N)}).\]  Just as with the leading factor $S^{(0)}$, it is possible to pull $S^{(1)}$ all the way to the front. Thus this ordering allows for a soft factor very reminiscent of the field theory result \[\sum_{\sigma\in \mathcal{P}_{N-3}} A_{{\rm YM}}(1,2_{\sigma},\ldots,(N-2)_{\sigma},q_{\sigma},N-1,N)F^{\sigma}_{(1,\ldots,N+1)}\to (S^{(0)}+S^{(1)})(A_{{\rm YM},N}F^{(2\ldots N-2)}_{(1\ldots N)})\] where $\mathcal{P}_{N-3}$ is the subgroup of the permutation group $S_{N-3}$ which keeps the order of $(2,3,\ldots,N-2)$ fixed and only moves $q$.
\newline

What about the other permutations $(2_{\sigma},3_{\sigma},\ldots,(N-2)_{\sigma},s_{\sigma})$? In fact, it is relatively easy to see what happens in these cases since most of the calculations are very similar to those presented in ssec.~\ref{sec:integr-appr-gener}. Firstly, it's helpful to consider subgroups of permutations $\sigma$ which preserve a definite ordering of the label $(2_{\sigma},\ldots,(N-2)_{\sigma})$ just as above and only move the soft particle $q$. Then there are essentially two cases to keep track of for each of these permutations:
\begin{enumerate}
\item The permutation $\sigma$ sends $N-2$ to a position in front of $s$. In this case there is always a leading contribution from the functions $F^{\sigma}(\alpha')$ equal to $1$ and a subleading contribution proportional to $s_{mq}$ ($m$ arbitrary) times logarithms. This will give a result very similar to \eqref{eq:approx}.
\item The permutation $\sigma$ sends $N-2$ to a position after $s$. The leading contribution in this case is a sum $\sum_{k=1}^{i}\frac{s_{k_{\sigma}q}}{z_{k_{\sigma}q}}$, where $i_{\sigma}$ indicates the position in front of the soft particle $(2_{\sigma},\ldots,i_{\sigma},s,(i+1)_{\sigma},\ldots,(N-2)_{\sigma})$. These results are very similar to \eqref{eq:result2} and \eqref{eq:result3}.
\end{enumerate}

After multiplying Weinberg's soft factor from the (appropriately permuted) field theory amplitudes and adding up all the contributions, we are once again led to the same contribution as in the identity permutation case. This concludes the proof that, in fact, 
\begin{align}
\mathcal{A}(1,2,\ldots,N-2,q,N-1,N) \to (S^{(0)} + S^{(1)})\mathcal{A}(1,2,\ldots,N-1,N)
\end{align} 
with $S^{(0)}$ given in \eqref{eq:knownfactor} and \[S^{(1)} = \frac{\epsilon_{\mu}q_{\nu}J^{\mu\nu}_{N-1}}{q.k_{N-1}} - \frac{\epsilon_{\mu}q_{\nu}J^{\mu\nu}_{N-2}}{q.k_{N-2}}\equiv S^{(1)}_{\rm YM}\]
in the soft limit of disk scattering amplitudes with an arbitrary number of legs and finite $\alpha'$ dependence. In short, there are no $\alpha'$-corrections to the subleading soft factor for open strings and the \emph{subleading soft factor is universal for tree-level string theory scattering amplitudes} as well. 

\subsection{Soft expansion of the 4pt disk amplitude}
\label{sec:soft-expansion-4pt}

Although we have seen the calculation for $N$ points, it might be instructive to work out the simplest cases in detail. As an example, consider the four point string theory amplitude. The four point disk amplitude can be written as
\[\mathcal{A}(1,2,3,4) = A_{\rm YM}(1,2,3,4)\frac{\Gamma(1+s)\Gamma(1+u)}{\Gamma(1+s+u)}\] where $s = \alpha'(k_{1} + k_{2})^{2} = \alpha'(k_{3}+k_{4})^{2}$ and $u = \alpha' (k_{1}+k_{4})^{2}$. Taking the soft particle to be $k_{4}$ and letting $k_{4}=q\to0$, there are two contributions. First of all, the Yang-Mills amplitude becomes the familiar \[A_{\rm YM}(1,2,3,4) \to S^{(0)}A_{\rm YM}(1,2,3)\] where $S^{(0)}$ is Weinberg's soft gluon factor
\[S^{(0)} = \frac{\epsilon.k_{1}}{q.k_{1}} - \frac{\epsilon.k_{3}}{q.k_{3}}.\] There is no subleading factor since it annihilates the three-particle amplitude \cite{Casali:2014xpa}. The soft expansion of Euler's Beta-Function
\[\frac{\Gamma(1+s)\Gamma(1+u)}{\Gamma(1+s+u)} = 1 -\alpha'^{2}\zeta_{2}s u + \ldots\] is the same as the $\alpha'$ expansion since all Mandelstam parameters depend on $q$. This means there is no subleading contribution in $q$ as the product $s u$ is $\mathcal{O}(q^{2})$. 
Collecting the lowest order contributions, we see that the four point string amplitude obeys the same soft relation as the field theory amplitude. It is of course well-known and trivial, but it is still amusing to point out that for the Beta-function the case of one soft string is ``identical'' to the field theory limit. 

\subsection{Soft expansion of the 5pt disk amplitude}
\label{sec:soft-expansion-5pt}

While the four point amplitude is essentially trivial, the five point amplitude is more interesting to our current discussion. It is given by the sum
\[\mathcal{A}_{5}(1,2,3,4,5) = A_{\rm YM}(1,2,3,4,5)F^{(23)}(\alpha') + A_{\rm YM}(1,3,2,4,5)F^{(32)}(\alpha').\] We take the particle $k_{3}$ (again with polarization vector $\epsilon$) to be the soft particle $k_{3}=q\to 0$ and shift all momenta greater than $3$ by minus one for later convenience. Then
\[\mathcal{A}_{5}(1,2,q,3,4) = A_{\rm YM}(1,2,q,3,4)F^{(2q)}(\alpha') + A_{\rm YM}(1,q,2,3,4)F^{(q2)}(\alpha').\] Both terms contribute to the same group of order-preserving permutations. Clearly, the field theory contributions are
\begin{align}
  A_{\rm YM}(1,2,q,3,4) &= (S^{(0)}_{3} + S^{(1)}_{3})A_{\rm YM}(1,2,3,4)\nln
  A_{\rm YM}(1,q,2,3,4) &= (S^{(0)}_{2} + S^{(1)}_{2})A_{\rm YM}(1,2,3,4)
\end{align}
and the contributions from the Euler integrals are given by
\begin{align}
\label{eq:fivepointexp}
F^{(23)} &\to B\star\lrbrk{1 + \sum_{i=1}^{2}s_{iq}\log z_{3,i} + s_{3q}\log z_{32}}\nln
F^{(32)} &\to B'\star\lrbrk{\frac{s_{12}}{z_{12}}s_{1q}\log\frac{z_{13}}{z_{12}} + \frac{s_{1q}}{z_{12}}}.
\end{align}
Since the second line in \eqref{eq:fivepointexp} is $\mathcal{O}(q)$ it doesn't contribute to the leading order.  Thus the leading order is given by the correct soft factor $S^{(0)}_{3} = \frac{\epsilon.k_{3}}{q.k_{3}}-\frac{\epsilon.k_{2}}{q.k_{2}}$. It remains to analyze the subleading factor. Notice that $B = B'\star\frac{s_{12}}{z_{12}}$, so that we can combine the two lines in \eqref{eq:fivepointexp} after pulling in the leading soft factors. After a quick examination we see that multiplication with Weinberg's soft factor $S^{(0)}_{3}$ in the limit of $F^{(2q)}$ and $S^{(0)}_{2}$ in the limit of $F^{(q2)}$ leads to the expected form and
\[\mathcal{A}(1,2,q,3,4) \to (S^{(0)}_{3} + S_{3}^{(1)})\mathcal{A}(1,2,3,4).\]

\section{Closed strings and soft factors from KLT}
\label{sec:closed-string-klt}

In this section we want to quickly suggest how to use the result presented in this work to find the subleading soft factors for the closed string (graviton) amplitudes. In the field theory limit, it is already known that the soft factors obey a double copy relation \cite{He:2014bga}. The existence of a double copy relation which relates ``squares'' of Yang-Mills amplitudes to gravity scattering amplitudes and a double copy relation for soft factors makes one speculate whether something similar might hold for string theory.

Of course we turned history upside down here. As a matter of fact, the double copy relations have their origin \cite{Berends:1988zp} in the \emph{Kawai-Lewellen-Tye (KLT) relations} \cite{Kawai:1985xq} between open string scattering amplitudes and closed string amplitudes.  The closed string amplitudes for graviton scattering can be derived from open string amplitudes in the left and right moving sectors with the help of these relations. In very condensed notation we can write these relations as \[\mathcal{M} = \mathcal{A}^{t} \mathcal{S} \mathcal{A}\] where $\mathcal{A}$ is a vector of the $(N-3)!$ independent color-orderings of the open string $N$ point amplitudes \cite{BjerrumBohr:2010hn}. $\mathcal{S}$ is a $(N-3)!\times (N-3)!$-matrix of additional sine-factors from the KLT relations and monodromy relations necessary to give the correct closed superstring amplitude \cite{Stieberger:2009hq,BjerrumBohr:2009rd}. All three factors are dependent on $\alpha'$. Remarkably, since $\mathcal{S}$ is a matrix of phase factors only dependent on the set of momenta $\{k_{i}\}$ and $\alpha'$, it follows from the open string amplitudes that the KLT relations are entirely general \cite{BjerrumBohr:2010hn}. They do not depend on the amount of supersymmetry, the type of compactification or the dimension $d$.
 
Since we have found that disk string scattering amplitudes behave in a way very much reminiscent of field theory scattering amplitudes, we can use this result, plug it into the KLT relations and extract the soft factors for genus zero closed string scattering amplitudes with massless (graviton) states. Notice that the right-moving sectors have slightly different soft limits from the one presented above and it will be necessary to use kinematic identities to derive the soft factors for graviton scattering. Also, with the current result we might not be able to derive the factor $S^{(2)}_{\rm g}$ since it might depend on higher order contributions from the Euler integrals.
\newline

Alternatively one could use the recently suggested CHY-like string scattering formula \cite{Bjerrum-Bohr:2014qwa} to find the soft factor for closed string scattering. The advantage of using this formula is that it mimics the form of the CHY formula \cite{Cachazo:2013iea,Cachazo:2013hca} which has already been used to find the subleading soft factor in arbitrary dimensions \cite{Schwab:2014xua}. One would expect that certain techniques that worked for the CHY formula might be applicable to the string theory formula.

Finally, it is also possible to ``ignore'' the result derived in this paper and go to the unintegrated form of the KLT relations found in more recent publications \cite{Schlotterer:2012ny} (see also \cite{Green:2013bza})
\[\mathcal{M}_{N} = \sum_{\sigma,\rho\in S_{N-3}} A^{\sigma}_{\rm YM} \mathcal{S}^{\sigma,\rho} \tilde{A}_{\rm YM}^{\rho}\] where the derivation of the subleading soft factors $S^{(i)}_{\rm g}$ would require an approximation of the $(N-3)!\times(N-3)!$-dimensional matrix 
\begin{align}
S^{\sigma,\rho} &= \int \prod_{j=2}^{N-2}d^{2}z_{j} \prod_{i<j}|z_{ij}|^{2s_{ij}}\lrbrk{\prod_{k=2}^{\lfloor N/2 \rfloor}\sum_{m=1}^{k-1}\frac{s_{m_{\sigma}k_{\sigma}}}{z_{m_{\sigma}k_{\sigma}}}\prod_{k=\lfloor N/2 \rfloor +1}^{N-2}\sum_{m=k+1}^{N-1}\frac{s_{m_{\sigma}k_{\sigma}}}{z_{m_{\sigma}k_{\sigma}}}}\nln &\qquad\qquad\qquad\times\lrbrk{\prod_{k=2}^{\lfloor N/2 \rfloor}\sum_{m=1}^{k-1}\frac{s_{m_{\rho}k_{\rho}}}{\bar{z}_{m_{\rho}k_{\rho}}}\prod_{k=\lfloor N/2 \rfloor +1}^{N-2}\sum_{m=k+1}^{N-1}\frac{s_{m_{\rho}k_{\rho}}}{\bar{z}_{m_{\rho}k_{\rho}}}}.
\end{align}
The obstacle here is the higher amount of poles that need to be treated in the approximation due to the unconstrained integration regions. The higher amount of poles is of course already a hint that the calculation \emph{can} give the expected graviton soft factors \eqref{eq:gravsoft}. We hope to cover the entire story for closed string amplitudes from KLT or one of the alternative approaches in an upcoming publication. 

\section*{Acknowledgments}
\label{sec:acknowledgments}

The author would like to thank Marcus Spradlin for suggesting the problem. I would further like to thank Steven Avery, Johannes Broedel, Robert de Mello Koch, Miguel Paulos, Matteo Rosso, and Anastasia Volovich for useful discussions. This work is supported by the US Department of Energy under contract DE-FG02-11ER41742.

\ifarxiv
\bibliographystyle{utphys}
\fi
\bibliography{soft}

\end{document}